\documentclass[Afour,sageh,times]{sagej}
\usepackage{moreverb,url}

\usepackage[colorlinks=true, linkcolor=blue, citecolor=blue, urlcolor=black]{hyperref}

\newcommand\BibTeX{{\rmfamily B\kern-.05em \textsc{i\kern-.025em b}\kern-.08em
T\kern-.1667em\lower.7ex\hbox{E}\kern-.125emX}}

\def\volumeyear{2024}

\usepackage{amsmath,amsfonts,bm}

\def\eqref#1{equation~\ref{#1}}
\def\1{\bm{1}}

\def\va{{\bm{a}}}

\def\vf{{\bm{f}}}

\def\vh{{\bm{h}}}

\def\vx{{\bm{x}}}

\def\vepsilon{{\bm{\epsilon}}}

\def\mA{{\bm{A}}}
\def\mB{{\bm{B}}}

\def\mE{{\bm{E}}}
\def\mF{{\bm{F}}}

\def\mH{{\bm{H}}}
\def\mI{{\bm{I}}}

\def\mL{{\bm{L}}}

\def\mR{{\bm{R}}}
\def\mS{{\bm{S}}}

\def\mX{{\bm{X}}}

\DeclareMathAlphabet{\mathsfit}{\encodingdefault}{\sfdefault}{m}{sl}
\SetMathAlphabet{\mathsfit}{bold}{\encodingdefault}{\sfdefault}{bx}{n}

\def\gL{{\mathcal{L}}}
\def\gM{{\mathcal{M}}}
\def\gN{{\mathcal{N}}}

\def\gP{{\mathcal{P}}}

\def\gU{{\mathcal{U}}}

\def\sR{{\mathbb{R}}}

\newcommand{\E}{\mathbb{E}}

\newcommand{\R}{\mathbb{R}}

\begin{document}

\runninghead{Lu etc.,}

\title{HPC-AI Coupling Methodology for Scientific Applications}

\author{Yutong Lu, Dan Huang and Pin Chen }

\affiliation{National Supercomputer Center in Guangzhou, School of Computer Science and Engineering, Sun Yat-sen University}
%\affilnum{2}SAGE Publications Ltd, UK}

\corrauth{Yutong Lu, Pin Chen}
%Alistair Smith, Sunrise Setting Ltd
%Brixham Laboratory,
%Freshwater Quarry,
%Brixham, Devon,
%TQ5~8BA, UK.}

\email{luyutong,chenp85@mail.sysu.edu.cn}

\begin{abstract}
Artificial intelligence (AI) technologies have fundamentally transformed numerical-based high-performance computing (HPC) applications with data-driven approaches and endeavored to address existing challenges, e.g. high computational intensity, in various scientific domains. In this study, we explore the scenarios of coupling HPC and AI (HPC-AI) in the context of emerging scientific applications, presenting a novel methodology that incorporates three patterns of coupling: surrogate, directive, and coordinate. Each pattern exemplifies a distinct coupling strategy, AI-driven prerequisite, and typical HPC-AI ensembles. Through case studies in materials science, we demonstrate the application and effectiveness of these patterns. The study highlights technical challenges, performance improvements, and implementation details, providing insight into promising perspectives of HPC-AI coupling. The proposed coupling patterns are applicable not only to materials science but also to other scientific domains, offering valuable guidance for future HPC-AI ensembles in scientific discovery.
\end{abstract}


\maketitle
\section{Introduction}
High-performance scientific and engineering computing has become an indispensable pillar that underpins scientific discovery, technology innovation, and large-scale engineering undertakings, positioning it as a critical element of national strategic capabilities~\cite{asch2018big}. Against this backdrop, recent successes of artificial intelligence (AI) in computer vision~\cite{liu2024sora} and natural language processing~\cite{achiam2023gpt} have catalyzed a growing convergence of AI methodologies with advanced computational frameworks. The trend of convergence has begun to reshape scientific and engineering computing paradigms: AI-driven approaches are increasingly deployed to accelerate large-scale simulations, optimize complex scientific workflows, and uncover intricate patterns within high-dimensional datasets~\cite{jumper2021highly,abramson2024accurate}. Taken together, these developments signal a powerful trend towards the integration of AI with high performance computing (HPC) systems, fostering a new era in which data-intensive, intelligent computational strategies drive both transformative scientific insights and the accelerated design of innovative engineering solutions.

As inquiries into multidisciplinary scientific and engineering problems are significantly growing in the aspects of complexity and scale, even reaching to extreme conditions, the technical challenges they present are escalating~\cite{dongarra2024co}. Conventional HPC strategies, while effective in traditional simulation paradigms, increasingly confront
formidable hurdles, including the soaring   computational costs of high-fidelity numerical methods for enhancing spatiotemporal resolution and significantly low efficiency of scaling simulations on next-generation supercomputing architectures~\cite{wang2024multi}. As an alternative, AI approaches, which have demonstrated remarkable efficiency and flexibility, remain constrained by limited training data availability, insufficient generalization capability, and challenges related to accuracy and interpretability~\cite{yang2024limits}.

By bridging the two methodologies, the predictive rigor of HPC-based physics-driven simulations with the adaptive and data-intensive AI models,  it offers a promising pathway towards a new computational research paradigm. By holistically combining HPC’s robust numerical foundations with AI`s heuristic efficiency, researchers can overcome the respective shortcomings of each approach. Then, a sophisticated HPC-AI methodology stands poised to become essential in tackling the complex and imperative scientific and engineering challenges. This aims to facilitate scalable, high-dimensional simulations, accelerate the solution of intricate physical models, and eventually build novel systems for more efficient scientific discovery. 

\begin{figure*}[htp]
    \centering
    \includegraphics[width=0.98\linewidth]{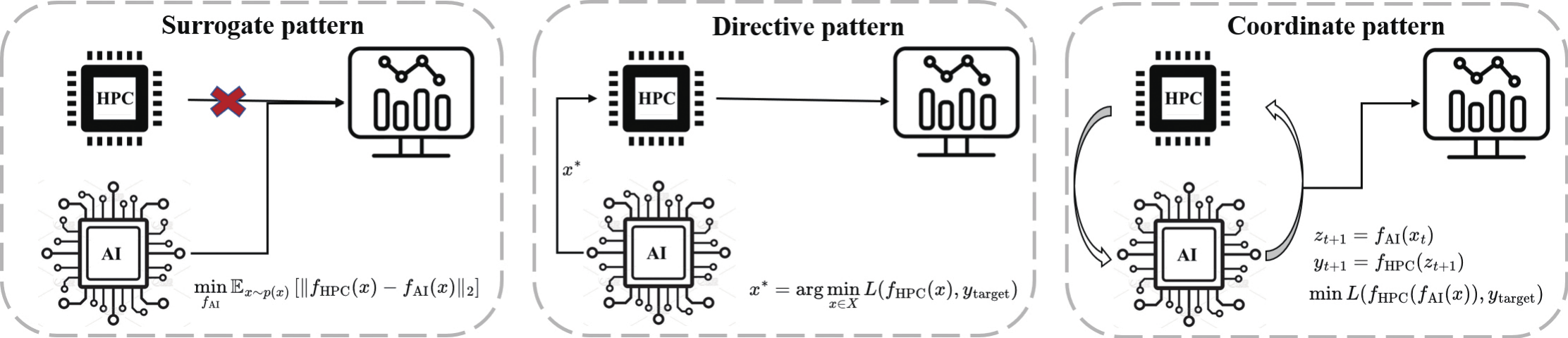}
    \caption{Illustrating the three patterns of HPC-AI coupling. (1) Surrogate pattern: AI models are used to replace part of entire simulations. (2) Directive pattern: AI provides real-time guidance to HPC simulations, optimizing parameters or providing intermediate feedback. (3) Coordinate pattern: HPC and AI operate interactively, exchanging data and feedback to solve complex problems in tandem.}
    \label{fig:arch}
\end{figure*}

In this study, we investigate the coupling of HPC and AI in order to overcome scientific discovery and engineering problems, proposing a methodology that encompasses three distinct coupling patterns: \textbf{ surrogate, directive and coordinate}. Each pattern represents a distinctive coupling strategy, e.g. AI-driven approximations that serve as surrogates for complex applications with HPC-AI ensembles that coordinate computing resources and tasks. Through comprehensive case studies in materials science, we develop patterns that exemplify each interaction pattern, highlight specific implementation strategies, and address the associated challenges. Our results demonstrate significant performance improvements and provide insights into the promising potentials of HPC-AI coupling. The proposed patterns not only advance materials science research but also are broadly applicable to other scientific domains, offering valuable guidance for future HPC-AI ensemble applications in the pursuit of scientific discovery.

The main contributions of this work are as follows:
\begin{enumerate} 
\item[1).] To achieve more efficient and effective integration of HPC and AI, we propose a novel methodology comprising three distinct coupling patterns: surrogate, directive, and coordinate.
\item[2).] We develop and enhance HPC-AI ensemble applications by exemplifying each interaction pattern in the comprehensive case studies, highlighting implementation strategies and addressing the technical challenges (e.g. limitations in accuracy and scalability). 
\item[3).] We demonstrate significant performance improvements in materials science applications, providing insights into the potential for future HPC-AI ensembles and offering applicable guidance for conducting HPC-AI driven researches across various scientific domains.
\end{enumerate}

\section{HPC-AI coupling Pattern}
In this section, we establish a fundamental methodology that  characterizes the variety of HPC-AI coupling patterns: \textbf{surrogate, directive, and coordinate}. We also provide a rigorous mathematical framework to facilitate their understanding and application. Although HPC and AI have independently spurred significant advances in various domains, their integration method remains obscure and ad hoc, lacking well-defined patterns for guiding HPC-AI ensemble research and development. By scrutinizing the key operational principles, data handling protocols, and computational workflows, we distill these diverse HPC-AI coupling practices into three overarching patterns, each of which represents a distinct form of convergence between HPC’s computational rigor and AI’s adaptive, learning-driven models as shown in Fig.~\ref{fig:arch}. Specifically, in the surrogate pattern, AI models supplant portions or the entirety of traditional HPC numerical simulations, thereby reducing computational overhead while expediting predictive workflows. In the directive pattern, AI provides real-time guidance to HPC processes, offering parameter optimization and intermediate feedback to steer simulations toward high-value solutions. Finally, the coordinate pattern establishes an interactive collaboration between HPC and AI, where both systems continuously exchange data and insights to address intricate scientific challenges in concert.
In the subsequent sections, we will present precise mathematical formulations to rigorously define these patterns and clarify the relationships among them. This structured approach lays the groundwork for a new generation of HPC-AI solutions that are with theoretical soundness and practical impact.

\subsection{Surrogate Pattern}

In the surrogate pattern, AI models are trained on existing simulation data to replace compute-intensive components of the simulation. By learning an approximate mapping from input parameters to target outputs, AI-driven models serve as efficient alternatives to complex physics-based simulations used in HPC, significantly reducing compute time.

\textbf{Definition:}

\begin{itemize}
    \item Let $x \in X$ represent the input parameters of the simulation, and $y \in Y$ denote the target outputs obtained from the HPC simulation.
    \item $f_{\text{HPC}}: X \rightarrow Y$ is the high-fidelity function employed in HPC simulations, typically computationally expensive.
    \item $f_{\text{AI}}: X \rightarrow Y$ is a data-driven surrogate model trained to approximate $f_{\text{HPC}}$ by minimizing the discrepancy between their outputs.
\end{itemize}

%The objective is to find $f_{\text{AI}}$ such that:
The objective is to find $f_{\text{AI}}$ that best approximates $f_{\text{HPC}}$ by minimizing a suitable discrepancy metric over the input distribution $p(x)$. For example, using the L2 norm, the optimization can be formulated as:
\[
\min_{f_{\text{AI}}} \mathbb{E}_{x \sim p(x)} \left[ \|f_{\text{HPC}}(x) - f_{\text{AI}}(x)\|_2 \right],
\]
Other norms or physics-informed loss functions can also be adopted depending on the nature of the simulation task.

The surrogate AI model $f_{\text{AI}}$ operates as a black-box function whose accuracy inherently depends on the quality and diversity of the training dataset $\{x_i, y_i\}$. It provides a computationally efficient approximation of $f_{\text{HPC}}$, facilitating rapid predictions without the need for extensive computational resources.

\subsection{Directive Pattern}

In the directive pattern, AI model works in conjunction with HPC by handling, configuring and optimizing the simulation tasks. It adjusts parameters or provides corrections based on intermediate simulation results. This model is particularly useful in high-dimensional parameter optimization problems, where AI can significantly reduce the search space and enhance computational efficiency.

\textbf{Definition:}

\begin{itemize}
    \item Let $X$ denote the input parameter space of the HPC simulation, and $f_{\text{HPC}}: X \rightarrow Y$ be the objective function of the HPC simulation.
    \item AI acts as an optimizer $f_{\text{AI}}$, dynamically adjusting the input $x$ to find the optimal input $x^*$.
    \item The optimization objective is defined as:
    \[
    x^* = \arg\min_{x \in X} L(f_{\text{HPC}}(x), y_{\text{target}}),
    \]
    where $L$ is a loss function measuring the discrepancy between $f_{\text{HPC}}(x)$ and the target output $y_{\text{target}}$.
\end{itemize}

The role of AI is to provide real-time guidance through a feedback mechanism, helping HPC simulations achieve better procedures and higher precision. By dynamically adjusting simulation parameters, AI enhances compute efficiency and accelerates convergence towards optimal solutions.

\subsection{Coordinate Pattern}

The coordinate pattern represents a multi-role coupling approach, where HPC and AI collaborate and incorporate third-party intelligent roles, such as pre-trained LLM and agents. Unlike directive pattern, AI module provides real-time insights to the HPC system, and the HPC system, in turn, supplies feedback to the AI models. In addition, the third-party roles can interact with AI module to provide external feedbacks. This multi-directional interaction allows for iterative refinement of AI-guided simulation inputs and HPC results through feedback exchange, leading to improved overall accuracy and performance. While the current framework does not explicitly involve continuous retraining of the AI model (as in active learning), it enables dynamic adaptation of simulation parameters based on real-time insights.

\textbf{Definition:}

\begin{itemize}
    \item Let $f_{\text{AI}}: X \rightarrow Z$ and $f_{\text{HPC}}: Z \rightarrow Y$ be the objective functions of the AI and HPC systems, respectively, where $Z$ represents the intermediate feedback information space.
    \item \textbf{Interaction Mechanism:} AI and HPC interact and update through an iterative feedback process, expressed as:
    \[
    z_{t+1} = f_{\text{AI}}(x_t), \quad y_{t+1} = f_{\text{HPC}}(z_{t+1}),
    \]
    where $x_t$ and $z_{t+1}$ represent the AI-guided simulation input parameters and the intermediate feedback information, respectively.
    \item \textbf{Objective:} Achieve optimal convergence through multiple iterations of interaction, defined as:
    \[
    \min L(f_{\text{HPC}}(f_{\text{AI}}(x)), y_{\text{target}}).
    \]
\end{itemize}

The interaction between AI and HPC occurs over multiple iterative steps, with continuous feedback exchange refining and optimizing the problem-solving process. This collaborative approach leverages the strengths of both AI and HPC, leading to improved compute efficiency and solution quality.

\section{HPC-AI Pattern Implementation}
In this section, we provide a concrete demonstration of the ensemble method of AI and numerical simulations within the domain of materials science, illustrating how these combined methodologies can drive scientific advances and engineering innovation. Specifically, we will detail the motivations that led to the formulation of these three coupling patterns and outline how each one can leverage AI`s pattern-recognition capabilities alongside the computational rigor of high-performance simulation tools. We will examine their core algorithmic components, data-flow protocols, and computational workflows, illustrating how subtle variations in the design strategy can lead to distinct advantages, including enhanced computational efficiency, improved AI model accuracy, better uncertainty quantification, or accelerated optimization of material properties.

To this end, we design three patterns as follows:
\begin{itemize}
  \item \textbf{Density Functional Theory calculation surrogate pattern} - Transformer AI models: A surrogate AI model that directly predicts material properties from structural information, bypassing the need to solve the Kohn-Sham (KS) equation.
  \item \textbf{Directive pattern} - Novel material structure space search: This pattern leverages a large-scale DFT-computed database to generate potentially novel material structures using the EquiCSP framework~\cite{LinICML24}. The generated structures are then evaluated through high-throughput DFT calculations using the Vienna ab initio simulation package (VASP)~\cite{wang1983density,chan2010efficient}.
  \item \textbf{Coordinate pattern for LLM-based material designing} - Agent-based interactive system for material design: An agent-driven interactive system that integrates material design workflows by utilizing large language models (LLMs) for decision-making. This system iteratively selects the most suitable tools, whether AI models or HPC computations, to efficiently optimize the design of target materials.
\end{itemize}
These patterns illustrate the complementary roles of the HPC and AI, providing a comprehensive method for accelerating modeling in materials science.

\subsection{Density Functional Theory Calculation Surrogate Pattern}

\begin{figure*}[t]
    \centering
    \includegraphics[width=0.8\linewidth]{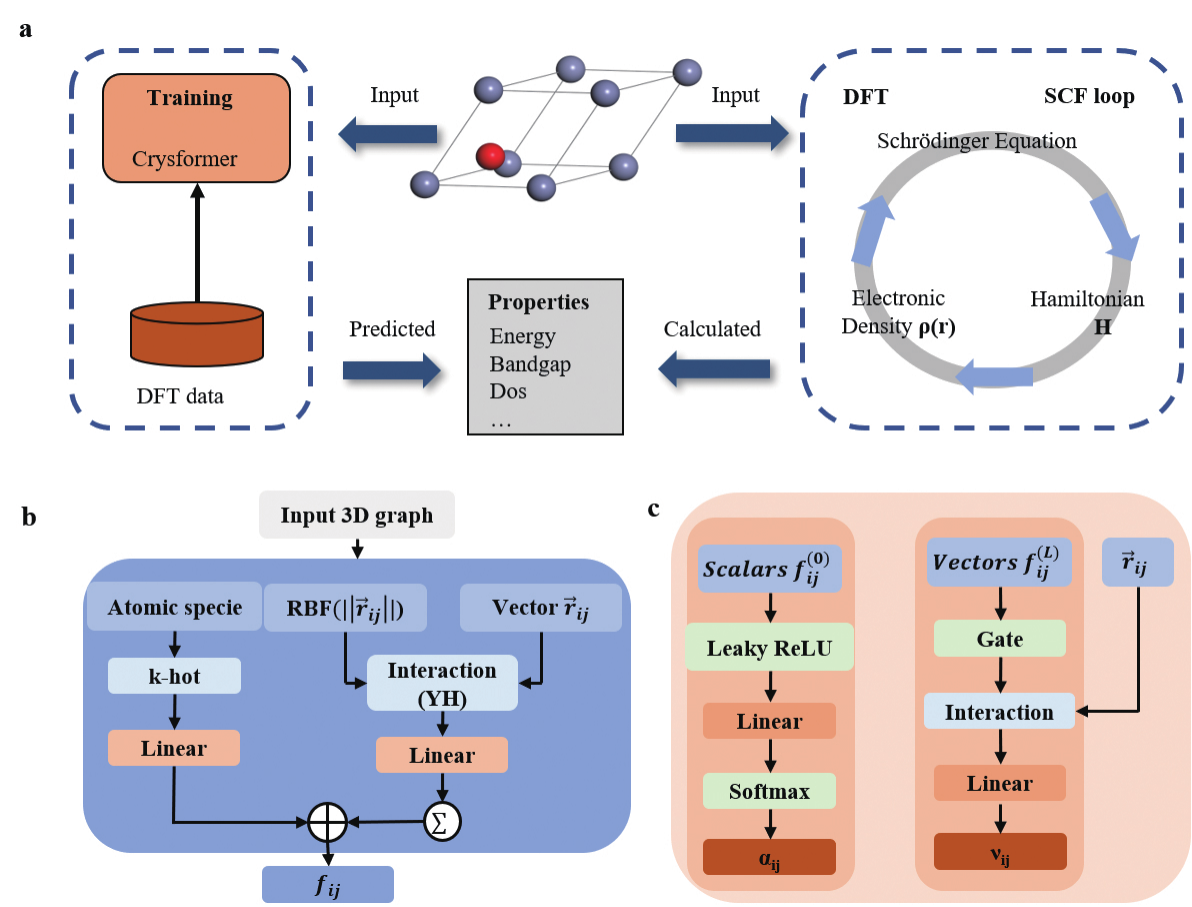}
    \caption{\textbf{a}. Schematic diagram of Crysformer replacing DFT for material property prediction. \textbf{b}. Methodology for 3D graph embedding method. \textbf{c}. Equivariant graph attention layer.
    }
    \label{fig:ai_model}
\end{figure*}

Density Functional Theory (DFT) is a quantum mechanical method widely used in materials science to calculate electronic structure and predict material properties from first principles~\cite{geerlings2003conceptual}. It provides a balance between accuracy and computational feasibility, making it a cornerstone in the simulation of solid-state systems. However, DFT remains computationally expensive, especially for large or complex systems, limiting its scalability in high-throughput applications. To overcome this limitation, AI models have been increasingly used to rapidly approximate DFT-level results through data-driven learning, enabling accelerated access to material property predictions~\cite{huang2023central}. In this context, DFT computations are often employed as the ground truth to train surrogate models that mimic their output.
As depicted in Fig.~\ref{fig:ai_model}a, DFT calculations typically involve solving the Schrödinger equation using iterative or approximate methods to derive key material characteristics, such as total energy and electronic band gaps. The computational complexity of these methods is influenced by various factors, including the size of the atomic system under consideration, the choice of functional, and the level of convergence accuracy required. Consequently, DFT simulations are often categorized as computationally intensive tasks. In contrast, AI approaches can directly infer material properties from the given structural inputs, effectively establishing a direct ``structure-property`` relationship without necessitating explicit solutions to the Schrödinger equation. By circumventing this iterative quantum-mechanical procedure, AI-driven models can substantially enhance computational efficiency and offer a promising alternative for rapid materials screening and discovery.

To estimate the properties of candidate crystal structures, we adopt a graph-based neural network model named Crysformer. This model aims to learn a mapping from atomic configuration to material properties by representing each crystal as a graph and applying geometric deep learning techniques. This prediction serves as a surrogate for DFT calculations during the screening process, significantly accelerating the evaluation of large numbers of generated structures.

Fig.~\ref{fig:ai_model}b and c provides a detailed illustration of the operational workflow in Crysformer for processing material structures. In this approach, the three-dimensional atomic coordinates of a given material are represented as vectorized input features, which are then fed into a neural network. During the training phase, the model parameters are iteratively refined to closely approximate the target material properties. Upon completion of training, Crysformer can directly infer these properties from structural data alone, eliminating the need for computationally demanding intermediate calculations. We will provide the details of Crysformer as follows.

\begin{enumerate}
    \item[1).] \textbf{Node embedding}. 
    In our graph network approach, the k-hot embedding technique~\cite{chen2022improving} is employed to construct the feature vector ($ \va_{i,k} $), effectively encoding the atomic properties of each atomic species.
    \item[2).] \textbf{Edge embedding}. 
    We proceed to investigate the 3D geometric properties, focusing on interatomic distances and the vectors $\vec{r}_{ij}$, through the application of spherical harmonics as described below:
   \begin{align}
   \label{eq:ele}
     \mE &=  {\mR\mB\mF}(\lVert \vec r_{ij} \rVert), \\
   \label{eq:edge}
     \vx_{ij} &= \varphi(\vh_{i}) + \varphi(\vh_{j}), \\
  \label{eq:edge_fea}
     \vf_{ij} &= \varphi_f(\vx_{ij} \otimes ^{TP}_{c\mE} \mS\mH (\vec r_{ij})) 
 \end{align}
 
In the given expression, ${\mR\mB\mF}(\lVert \vec{r}_{ij} \rVert)$ denotes the Radial Basis Function (RBF) expansion for the distance between atoms. The initial edges are constructed using the k-nearest neighbor (kNN) method, as described in~\cite{Yan2022PeriodicGT}. The feature vector $\vx_{ij}$ combines the attributes of node $i$ (target) and node $j$ (source) through linear layers to form the initial message. The term $\mS\mH(\vec{r}_{ij})$ represents the spherical harmonics (SH) embeddings~\cite{gasteiger2020directional} of the relative position $\vec{r}_{ij}$, and $c \mE$ is a weight parameterized by $\mE$. Finally, the non-linear messages and attention weights are computed via $\vf_{ij}$.

\end{enumerate}
\subsubsection{Equivariant Graph Attention}
Given $\vf_{ij}$, which contains multiple type-$L$ vectors representing SE(3)-equivariant irreducible representations (irreps) features, we partition $\vf_{ij}$ into two components: $\vf_{ij}^{0}$ and $\vf_{ij}^{L}$. The scalar component $\vf_{ij}^{0}$ is invariant under input transformations, whereas $\vf_{ij}^{L}$ consists of type-$L$ vectors, which can disrupt equivariance. Following the methodology outlined in~\cite{liao2023equiformer}, we apply distinct operations to each subset of $\vf_{ij}$.

\begin{figure*}[htp]
    \centering
    \includegraphics[width=0.95\textwidth]{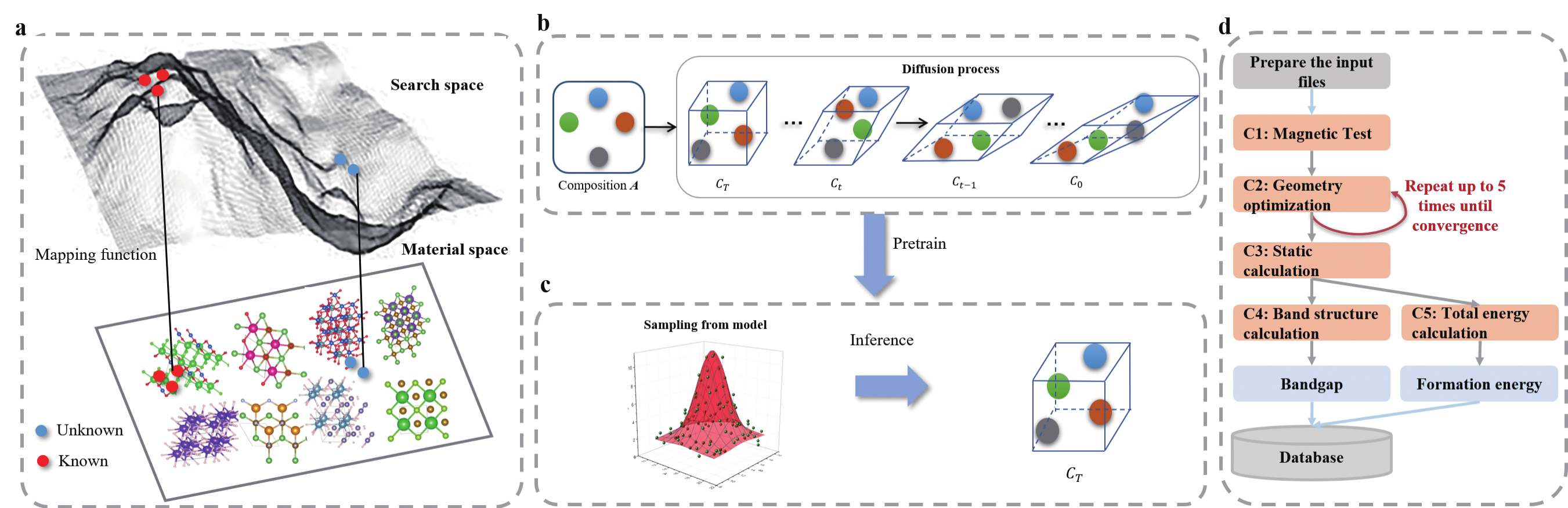}
    \caption{
    \textbf{a}. A projected 3D material structure search space.
\textbf{b}. Schematic of the crystal diffusion model.
\textbf{c}. Schematic of crystal ab initio generation.
\textbf{d}. DFT calculation workflow of the generated material structure.
    }
    \label{fig:mode2}
\end{figure*}

\begin{enumerate}
    \item[1).] \textbf{Type-0 features.}
    For $\vf_{ij}^{0}$, we apply the leaky ReLU activation function, followed by a softmax operation, to compute $\alpha_{ij}$:
    \begin{align}
        \zeta_{ij} &= \alpha^{\top} LeakReLU(\vf_{ij}^{0}) , \\
        \alpha_{ij} &= \frac{exp(\zeta_{ij})}{\sum_{k\in\gN(i)}exp(\zeta_{ik})}
    \end{align} 
    Here, $\alpha$ is a trainable vector with the same dimensionality as $\vf_{ij}^{0}$, and $\zeta_{ij}$ is a scalar value.
    \item[2).] \textbf{Type-L features.} 
    A non-linear transformation is applied to $\vf_{ij}^{L}$ to produce a non-linear message:
    \begin{align}
        \mu_{ij} &= Gate(\vf_{ij}^{L}), \\
        \upsilon_{ij} &= \varphi_f(\mu_{ij} \otimes ^{TP}_{\omega} \mS\mH (\vec r_{ij}))
    \end{align}
    We utilize the equivariant gate activation, as outlined in~\cite{weiler20183d}, to modulate the output features in a symmetry-preserving manner. This mechanism allows scalar and tensorial features to interact while maintaining SE(3) equivariance. Subsequently, a method similar to that described in Eq.~\ref{eq:edge_fea} is employed to compute the message $\upsilon_{ij}$ passed from node $j$ to node $i$, integrating atomic features and geometric relationships through gated non-linear transformations.
\end{enumerate}

In the final step, $\alpha_{ij}$ and $\upsilon_{ij}$ are converted into scalars via multiplication. A mean aggregation is then performed across all nodes to predict the property value.
\begin{align}
    \gP (i) &= \frac{1}{|\mathcal{N}(i)|} \sum_{j \in \mathcal{N}(i)} \alpha_{ij} \cdot \upsilon_{ij} , \\
    \gP &=  \frac{1}{|\mathcal{V}|} \sum_{i \in \mathcal{V}} T_{c}(i) 
\end{align}

\subsection{Directive Pattern for Materials Structure Space Search}
In materials discovery, a central task is the search for stable and synthesizable crystal structures across vast chemical and structural spaces. This problem, often referred to as crystal structure prediction (CSP), is challenging due to the combinatorial explosion of possible atomic arrangements, compositions, and symmetries~\cite{woodley2008crystal}. Traditional search algorithms typically rely on heuristic strategies and stochastic sampling, which require extensive DFT evaluations to identify energetically favorable structures.
As shown in Fig.~\ref{fig:mode2}, exploring the compound space in materials design represents a significant challenge, especially in the discovery of novel materials. Traditional approaches, such as genetic algorithms~\cite{2006Crystal,GLASS2006713}, particle swarm optimization~\cite{call2007global}, or random search~\cite{pickard2011ab}, are computationally expensive, thereby limiting the depth and breadth of the exploration. We designed a diffusion generative model that characterizes known structures as latent variables, from which element types, crystal cell structures, and atomic coordinates are sampled to generate new structures. These newly generated structures are then input into DFT calculation software for structural optimization and subsequent property calculations.

EquiCSP, introduced in our previous work~\citep{LinICML24}, is a diffusion-based framework specifically designed to learn stable structure distributions for the Crystal Structure Prediction (CSP) task. This method leverages a periodic E(3) equivariant model, enabling the joint optimization of the lattice matrix $\mL$ and fractional coordinates $\mF = \mL^{-1}\mX$. A comprehensive description of the methodology can be found in the original publication. Building on the EquiCSP framework, we have developed an advanced architecture that significantly enhances its capability for constructing large-scale materials databases.

The composition $\mA = [\va_1, \va_2, \ldots, \va_N]$ plays a critical role in constructing a database with a uniform distribution of elements $a_i$ and consistent structure sizes $N$ within a unit cell. In the EquiCSP method, the composition $\mA$ is treated as a continuous variable in real space $\sR^{h \times N}$, allowing for the utilization of the standard DDPM-based approach~\cite{hoogeboom2022equivariant}. The forward diffusion process is defined as:

\begin{align}
\label{eq:la0}
q(\mA_t | \mA_{0}) &= \gN\Big(\mL_t | \sqrt{\bar{\alpha}_t}\mA_{0}, (1 - \bar{\alpha}_t)\mI\Big). 
\end{align}

Here, the variance is modulated by $\beta_t \in (0,1)$, with $\bar{\alpha}_t = \prod_{s=1}^t \alpha_t = \prod_{s=1}^t (1 - \beta_t)$, following the cosine scheduler~\cite{nichol2021improved}. The backward generation process is expressed as:

\begin{align}
p(\mA_{t-1} | \gM_t) &= \gN(\mA_{t-1} | \mu_\mA(\gM_t),\sigma_\mA^2 (\gM_t)\mI), 
\end{align}
where $\mu_\mA(\gM_t) = \frac{1}{\sqrt{\alpha_t}}\Big(\mA_t - \frac{\beta_t}{\sqrt{1-\bar{\alpha}_t}} \hat{\vepsilon}_\mA(\gM_t, t)\Big)$ and $\sigma_\mA^2(\gM_t) = \beta_t \frac{1-\bar{\alpha}_{t-1}}{1-\bar{\alpha}_t}$. The model predicts the denoising term $\hat{\vepsilon}_\mA(\gM_t, t) \in \R^{h \times N}$. Training is aimed at minimizing the one-hot diffusion loss:

\begin{align}
    \gL_{\mA,\text{continuous}} &=\E_{\vepsilon_\mA\sim\gN(0,\mI), t\sim\gU(1,T)}[\|\vepsilon_\mA - \hat{\vepsilon}_\mA(\gM_t,t)\|_2^2].
\end{align}

The combined training objective for the joint diffusion model, encompassing $\mL$, $\mF$, and $\mA$, is defined as follows:
\begin{align}
    \gL_\gM &= \lambda_\mL \gL_\mL + \lambda_\mF \gL_\mF + \lambda_\mA \gL_\mA.
\end{align}
For the CSP task, we set $\lambda_\mL = \lambda_\mF = 1$ and $\lambda_\mA = 0$, as $\mA$ remains constant during generation. In contrast, for the \textit{ab initio} generation task, $\lambda_\mL$ and $\lambda_\mF$ are kept at 1, while $\lambda_\mA$ is assigned a larger value to balance the scales of the loss components.

\begin{figure*}[htp]
    \centering
    \includegraphics[width=0.95\textwidth]{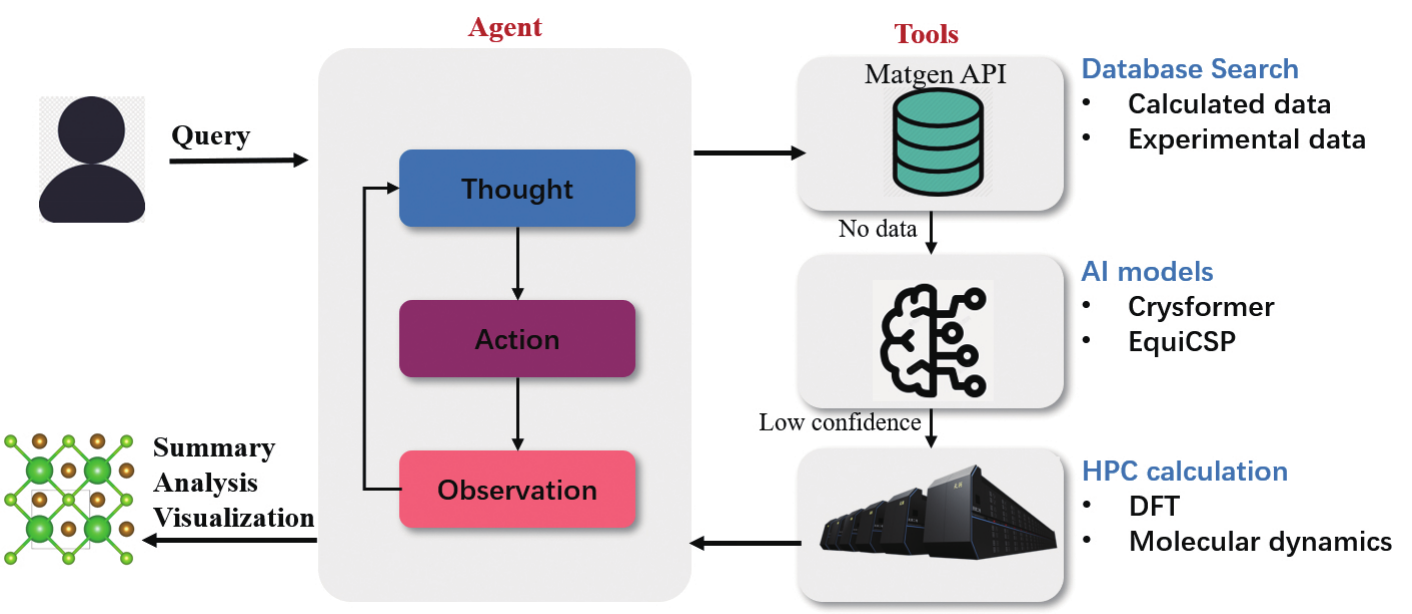}
    \caption{Hierarchical ReAct agent planning in coordinate pattern. Deployment via a standardized langchain interface with support for hierarchical tool invocation, including material data repository, DFT calculation workflows and AI pre-trained models.}
    \label{fig:model3}
\end{figure*}

Initially, we faced the challenge of obtaining a sufficient number of effective components $\mA$ for structure generation via the CSP method. To address this, we extract $\mA$ from the hidden space distribution of a pre-trained model, enabling the rapid generation of a large number of structures. Subsequently, we employ the CSP technique to focus on generating structures corresponding to the underrepresented $\mA$ within our database. The detailed methodology is as follows:

\begin{enumerate}
    \item[1).] \textbf{Extensive sampling by Ab initio method}. 
    This approach involves predicting structures \textit{de novo} by leveraging a predefined atomic distribution from the pre-trained model. In our process, since the number of atoms per unit cell $N$ remains constant during generation, we first select $N$ based on its distribution in the training set, following the method outlined in ~\citep{hoogeboom2022equivariant}. This results in a sampled distribution expressed as $p(\gM, N) = p(N)p(\gM \mid N)$, where $p(N)$ is derived from a pre-computed data distribution, and $p(\gM \mid N)$ is effectively modeled using EquiCSP.
    \item[2).] \textbf{Fine-tuning data by CSP generation}.
    A CSP approach is employed to fine-tune the elemental distribution in the database, specifically by utilizing the element $\mA$ as an input parameter for sampling.
\end{enumerate}

\textbf{Pre-Training}. 
We pre-train the model on approximately 1.14 million non-redundant 3D crystal structures obtained from existing databases, including the Materials Project\footnote{\url{https://next-gen.materialsproject.org}}, OQMD\footnote{\url{https://www.oqmd.org}}, Matgen\footnote{\url{https://matgen.nscc-gz.cn}}, and ICSD.

\subsection{Coordinate Pattern for LLM-based Material Designing}

In the context of HPC-AI collaborative scheduling, we propose a streamlined workflow for materials design, leveraging the complementary strengths of AI predictions and DFT computations. For material properties of interest, the process begins with AI-based predictions, which are rapid but may lack precision. To address this, we incorporate a confidence evaluation mechanism where the AI-generated results are assessed for reliability. If the confidence is regarded as insufficient, a more accurate yet time-intensive DFT calculation is triggered.

As shown in Fig.~\ref{fig:model3}, this workflow is implemented through an intelligent agent that dynamically coordinates between AI and DFT, optimizing the computational resources of HPC systems. Over time, the DFT results obtained from low-confidence predictions are buffered, and once a threshold (e.g., in sample size or time interval) is reached, the AI model undergoes periodic fine-tuning using these high-fidelity data. This asynchronous refinement allows the model to gradually improve its predictive performance without interrupting the ongoing inference process.
This iterative refinement establishes a positive feedback loop, improving AI reliability while reducing dependency on exhaustive DFT calculations. Such a synergistic approach ensures efficient utilization of HPC resources, accelerates the material discovery process, and continuously refines the AI's capability to predict material properties with greater precision, paving the way for scalable and intelligent materials design frameworks.

In this pattern, we integrate a decision-making agent that interacts with both a pre-computed DFT database and AI-based predictive models to provide material property predictions and generate novel structures if necessary. The procedure is as follows:

1). \textbf{Database Query:} Given a user-provided material with input representation $\mathbf{X}$ (e.g., atomic coordinates, lattice parameters, composition), the agent first searches a DFT calculation database:
   \[
   \text{Search}(\mathbf{X}) \rightarrow \mathcal{D}_{\text{DFT}}.
   \]
   If a matching structure and its computed properties are found, the agent immediately returns the known properties.

2). \textbf{AI-Based Prediction or Generation:} If no match is found in the database, the agent resorts to AI models:
   \[
   \hat{y} = f_\theta(\mathbf{X}),
   \]
   where $f_\theta$ represents an AI model (e.g., a Transformer-based model for property prediction or a diffusion-based generative model for novel structure generation).

3). \textbf{Confidence Estimation for Property Prediction:} For property prediction using Transformer-based models, we estimate confidence through an uncertainty quantification technique. Suppose we employ Monte Carlo (MC) Dropout or an ensemble approach to obtain a set of predictions $\{\hat{y}^{(1)}, \hat{y}^{(2)}, \dots, \hat{y}^{(T)}\}$ for the same input $\mathbf{X}$. The predictive mean and variance are computed as:
   \[
   \hat{\mu} = \frac{1}{T} \sum_{t=1}^{T}\hat{y}^{(t)}, \quad
   \hat{\sigma}^2 = \frac{1}{T}\sum_{t=1}^{T}(\hat{y}^{(t)} - \hat{\mu})^2.
   \]
   A small variance $\hat{\sigma}^2$ indicates high confidence, and thus the model’s prediction $\hat{\mu}$ is considered reliable.

4). \textbf{Confidence Estimation for Structure Generation:} For structure generation (e.g., using a diffusion-based crystal structure prediction model, EquiCSP), we assess confidence with a Match Discriminator (MD). Inspired by Diffdock~\cite{CorsoSJBJ23}, we first generate multiple samples for each composition:
   \[
   \{ \mathbf{X}^{(1)}, \mathbf{X}^{(2)}, \dots, \mathbf{X}^{(5)} \}
   \]
   from the diffusion model. Let $\mathbf{X}_{\text{gt}}$ be the ground-truth structure. We compute the Root Mean Square Deviation (RMSD) for each generated structure:
   \[
   \text{RMSD}^{(i)} = \sqrt{\frac{1}{N}\sum_{n=1}^{N} \| \mathbf{r}_n^{(i)} - \mathbf{r}_n^{(\text{gt})} \|^2 },
   \]
   where $\mathbf{r}_n^{(i)}$ and $\mathbf{r}_n^{(\text{gt})}$ denote atomic positions of the $n$-th atom in the $i$-th generated structure and the ground truth, respectively. We label each sample as a match if $\text{RMSD}^{(i)} < d$, where $d$ is a predetermined threshold.

   A binary classifier (the MD) is trained to predict the probability $p_{\text{match}}$ that a newly generated structure is a “match”:
   \[
   p_{\text{match}} = g_\phi(\mathbf{X}), 
   \]
   where $g_\phi$ is the MD`s predictive function. A high $p_{\text{match}}$ indicates that the generated structure is likely close to the ground truth distribution, thereby serving as a confidence score for generative predictions.

5. \textbf{Decision Criterion:} The agent then sets a confidence threshold $\tau$. If the property prediction variance $\hat{\sigma}^2$ is sufficiently low (for property prediction models) or if $p_{\text{match}}$ is sufficiently high (for structure generation), i.e.,
   \[
   \hat{\sigma}^2 < \tau \quad \text{or} \quad p_{\text{match}} > \tau,
   \]
   the predicted or generated result is deemed trustworthy and is returned to the user.

   Conversely, if the confidence is low, the agent initiates direct DFT calculations:
   \[
   \text{DFT}(\mathbf{X}) \rightarrow \hat{y}_{\text{DFT}}
   \]
   involving structure optimization and property computation from first principles. While computationally more expensive, this ensures reliable outcomes when AI-based predictions or generated structures do not meet the required confidence criteria.

By integrating the agent with both uncertainty-quantified prediction models and a match discriminator for structure generation, we achieve a robust and adaptive system that dynamically chooses between leveraging existing databases, employing AI model outputs, or resorting to computationally intensive DFT calculations based on estimated confidence.

\section{Experiments and Results}
In this section, we conduct experiments on three designed HPC-AI coupling patterns. All experiments are performed on highend HPC resources. The AI model training and inference are carried out on NVIDIA A800 GPUs. For DFT calculations, we utilize the VASP version 5.4.4~\cite{wang1983density,chan2010efficient}, running on Intel(R) Xeon(R) Platinum 8358P processors.

\subsection{Surrogate Pattern for DFT Calculation}

\begin{figure}
    \centering
    \includegraphics[width=0.35\textwidth]{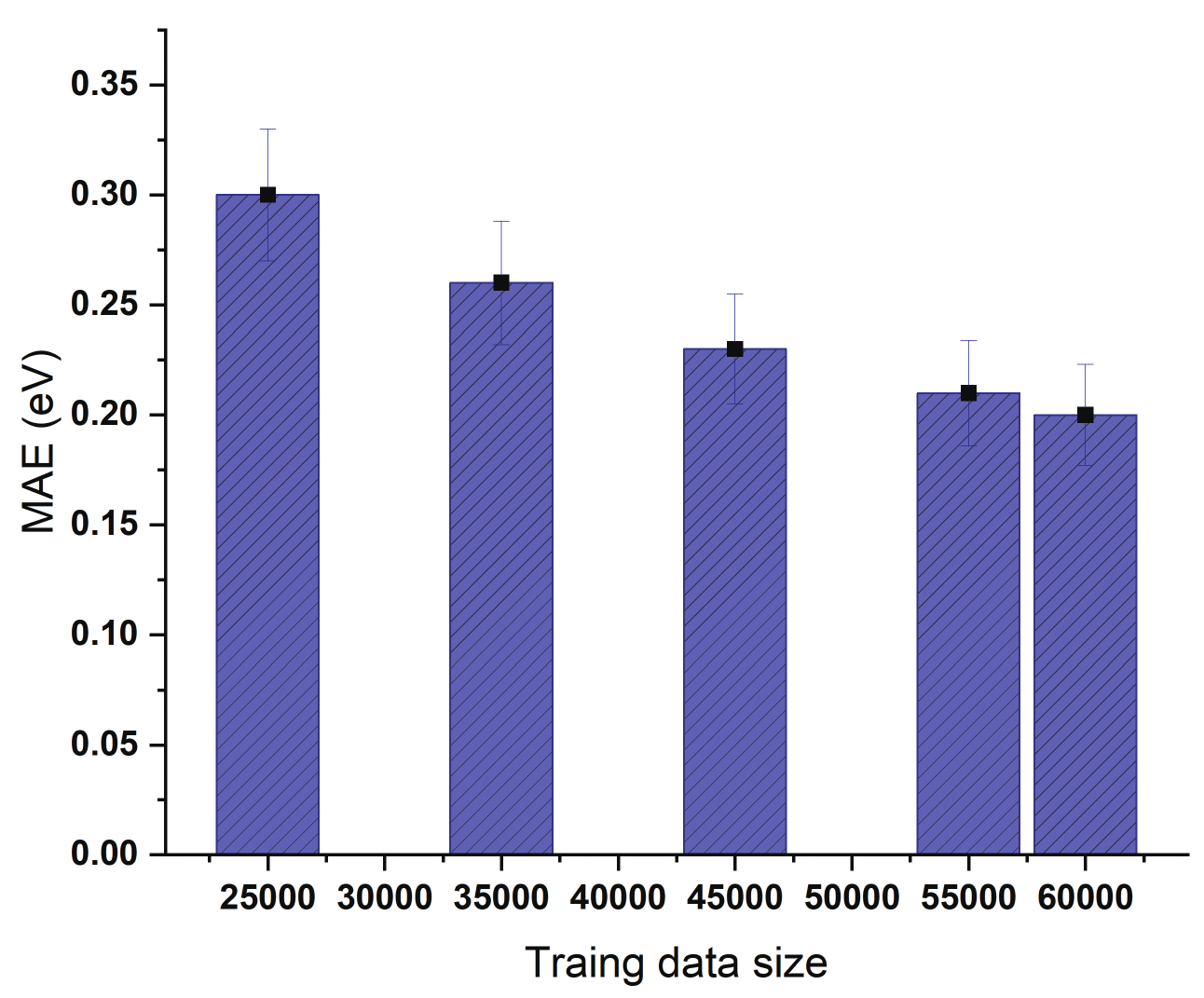}
    \caption{Impact of dataset size on MAE performance for bandgap property prediction.}
    \label{fig:data_size}
\end{figure}

We use the MP dataset, consisting of 69,239 crystal structures, which has been widely employed for training other GNN models, to evaluate Crysformer. As shown in Table~\ref{tab:ai_performance}, our Crysformer achieves state-of-the-art performance in predicting both formation energy and bandgap. Notably, Crysformer slightly outperforms the second-best GNN model, Matformer~\cite{Yan2022PeriodicGT}, another transformer-based approach. This highlights Crysformer's superior ability to capture 3D geometric information through edge vector representations.

\begin{table}[htp]
  \centering
  \caption{
  Comparative AI-based models for property prediction. 
  }
  \label{tab:ai_performance}
  \begin{tabular}{lccc}
    \toprule
    & \multicolumn{3}{c}{\textbf{Material Project}}\\
    %\cline{2-3} \cline{3-4}
    \cmidrule{2-3} \cmidrule{3-4}
    & & \textbf{Formation energy} & \textbf{Bandgap} \\
    %\cline{3-3} \cline{4-4}
    \cmidrule{3-3} \cmidrule{4-4}
    \textbf{Method}  &\textbf{Data size} &E (eV/atom)  & E$_g$ (eV)\\
    \midrule
    CGCNN & 16,485    &0.039 &0.388\\
    MEGNet & 36,720 &0.028 &0.33\\
    CrystalNet & 60,000  &0.030 &0.285\\
    ALIGNN & 60,000 &0.022 &0.218\\
    Matformer & 60,000 &0.021 & 0.211 \\
    \midrule
    Crysformer & 60,000 &\textbf{0.020} &\textbf{0.207} \\ 
  \bottomrule
\end{tabular}
\end{table}

We further evaluate the impact of data size on the performance of the Crysformer model, as illustrated in Fig.~\ref{fig:data_size}. The MAE decreases from 0.31 eV to 0.207 eV as the training sample size increases from 25,000 to 60,000. Training sizes were selected at intervals of 5,000 to capture the overall trend while maintaining computational feasibility. This demonstrates that data-driven AI models can significantly benefit from the construction of larger data repositories.

\begin{table}[ht]
\caption{Performance comparison of ML and DFT calculated models on selected data. Reference data obtained from ~\cite{choudhary2018computational}, \cite{chenmegnet2019}
}
\centering
\begin{tabular}{lcc}
\hline
\textbf{Methods} & \textbf{MAE (eV)} & \textbf{Time per structure} \\
\hline
\textbf{ML Models} & - & (s) \\
Crysformer & 1.18 & 1.66 \\
Crysformer-TL & 0.70 & 1.57 \\
\hline
\textbf{PBE-Based DFT} &- & (min) \\
MP & 1.38 & - \\
Matgen & 1.21 & 24.5 \\
AFLOW & 1.20 & - \\
OQMD & 1.09 & - \\
\hline
\textbf{HSE-Based DFT} &- & (min)\\
HSE & 0.41 & 228.1 \\
\hline
\end{tabular}
\label{tab:wet}

\end{table}

Specifically, following the method described in~\citep{choudhary2018computational}, we conducted tests as shown in Table~\ref{tab:wet}. For bandgap prediction using the PBE functional dataset, Crysformer achieved a prediction accuracy comparable to DFT calculations with the PBE functional, while being approximately 886 times more efficient. For the high-accuracy DFT results obtained with the HSE functional, the performance of Crysformer-TL was slightly lower than that of DFT calculations, yet it achieved an efficiency improvement of approximately 8,245 times. These results indicate that surrogate patterns are limited by the quality of the training data in terms of prediction accuracy~\cite{xie2018crystal,chen2021learning}. However, the predicted results closely match those of DFT calculations, and the computational efficiency is vastly superior to that of DFT.

\subsection{Directive Pattern for Materials Structure Space Search }  

\begin{figure}
    \centering
    \includegraphics[width=0.5\textwidth]{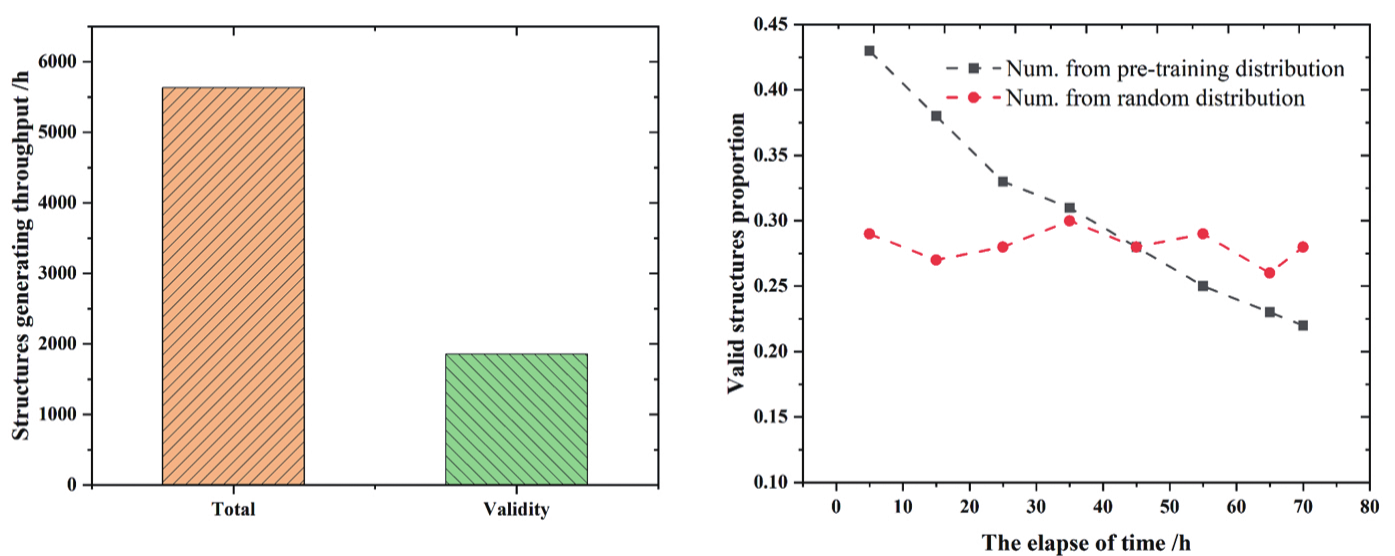}
    \caption{
Experimental Results.
\textbf{a}. Throughput of structure generation using DiffCSP.
\textbf{b}. Evolution of the proportion of valid structures over time.
    }
    \label{fig:dire_thr}
\end{figure}

\begin{figure}
    \centering
    \includegraphics[width=0.4\textwidth]{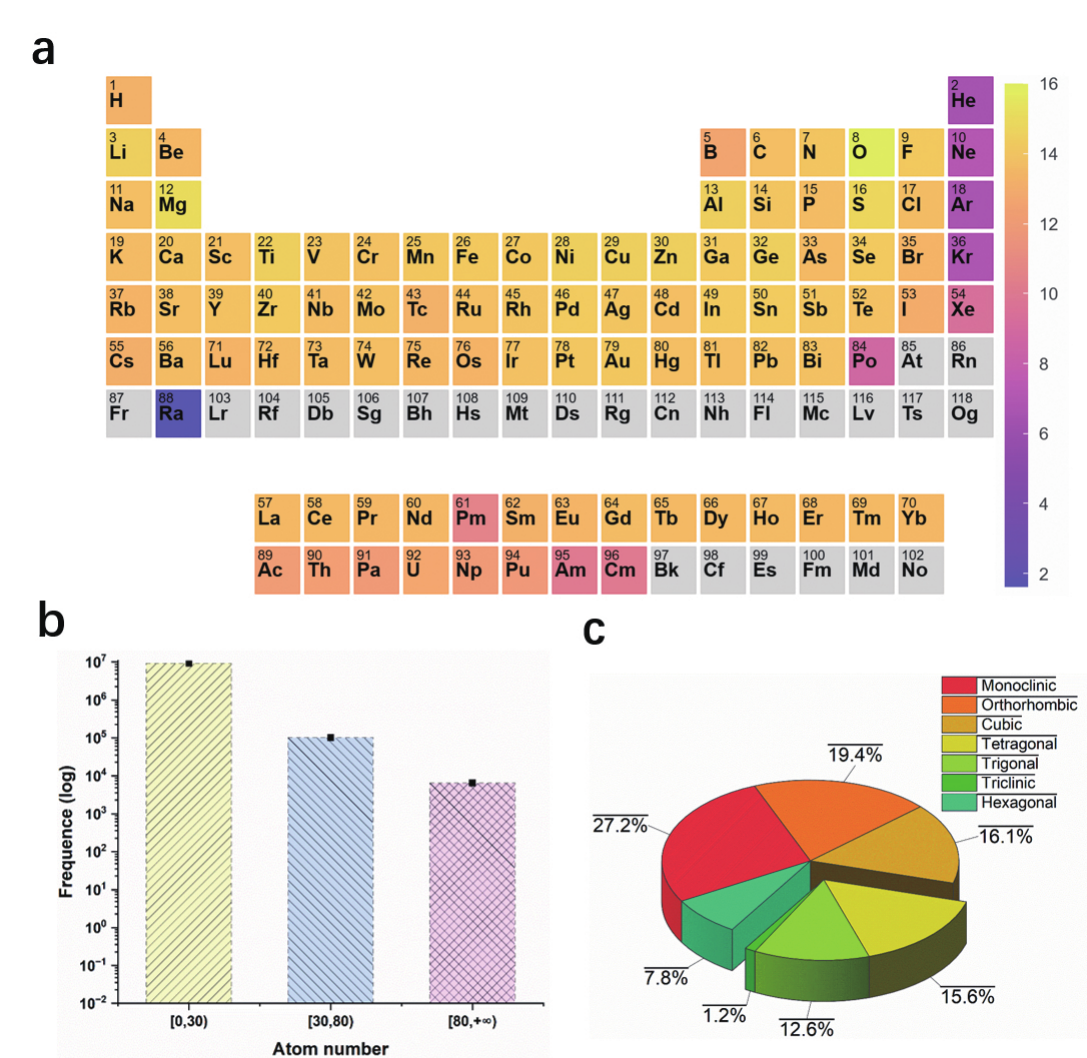}
    \caption{Statistics of Material Structures Generated by Generative Models.
\textbf{a}. Frequency distribution of chemical species.
\textbf{b}. Distribution of atom counts in primary unit cells.
\textbf{c}. Classification distribution across crystal systems.
}
    \label{fig:dire_gen}
\end{figure}

\begin{figure}
    \centering
    \includegraphics[width=0.48\textwidth]{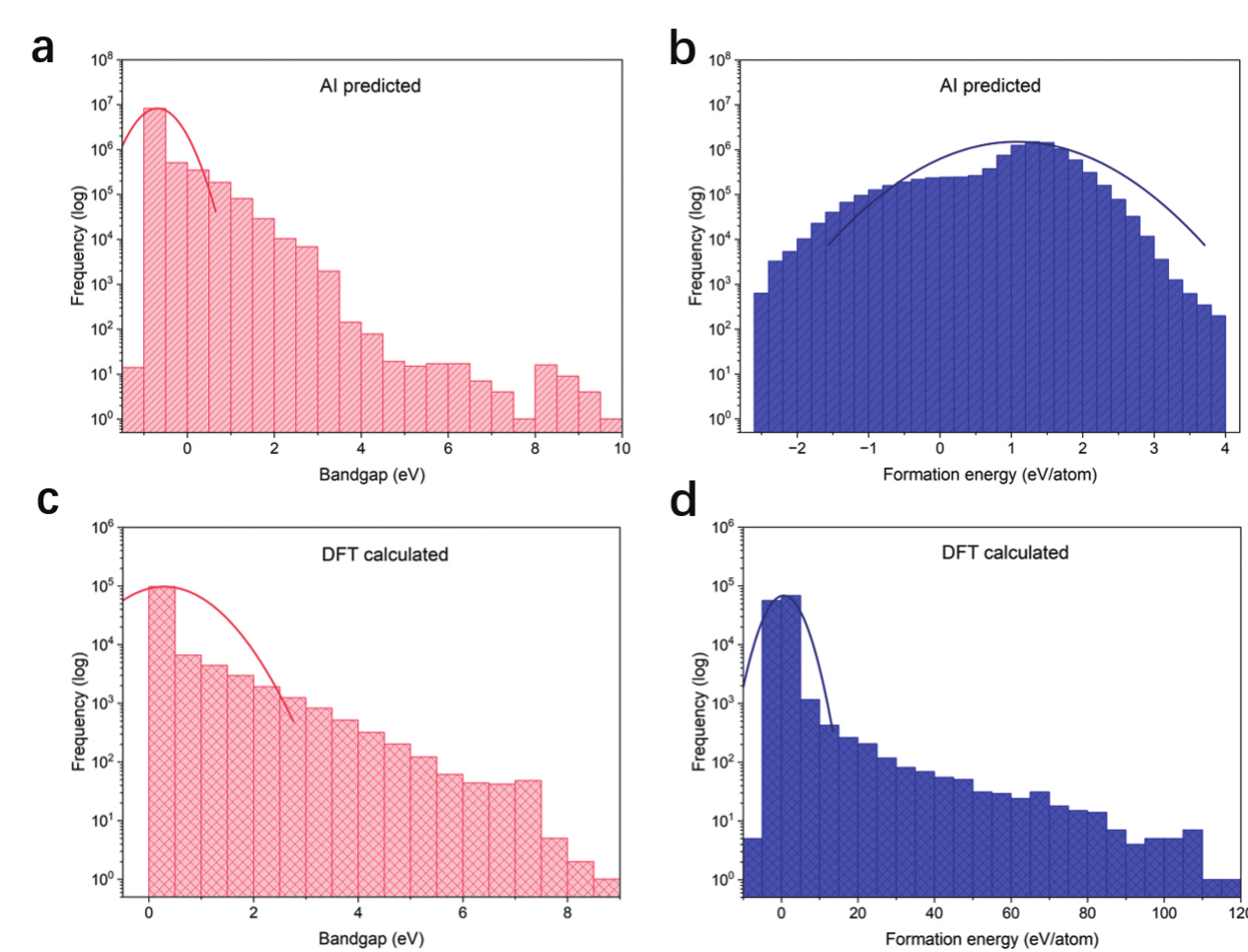}
    \caption{
    Frequency distribution of predicted bandgap \textbf{a} and formation energy values \textbf{b}, respectively.  Frequency distribution of DFT calculated bandgap \textbf{c} and formation energy \textbf{d} values, respectively.
    }
    \label{fig:prop_dist}
\end{figure}

As illustrated in Fig.~\ref{fig:dire_thr}a, we utilize EquiCSP to perform de novo sampling for large-scale structure generation. This approach enables the generation of approximately 5,663 structures per hour, among which around 1,859 are preliminarily validated on an hourly basis. For structure generation, a preliminary validity assessment is conducted using the following methods to ensure the quality of the sampled structures.

\begin{enumerate}
    \item \textbf{Reasonable chemical compositions}: We remove the structure with number of chemical components that is more than 10.
    \item  \textbf{Electroneutrality approach}: We calculate the oxidation states of each element using the SCMAT toolkit~\cite{davies2019smact} and remove structures with charge imbalance.
    \item  \textbf{Exhibiting symmetry}: Observing that materials in nature exhibit high symmetry, structures with space group equal to 1 will be excluded.
    \item  \textbf{Structural similarity}. A structural similarity algorithm is employed to further eliminate structures that are either similar or identical.
\end{enumerate}

\begin{figure*}[htp]
    \centering
    \includegraphics[width=0.95\textwidth]{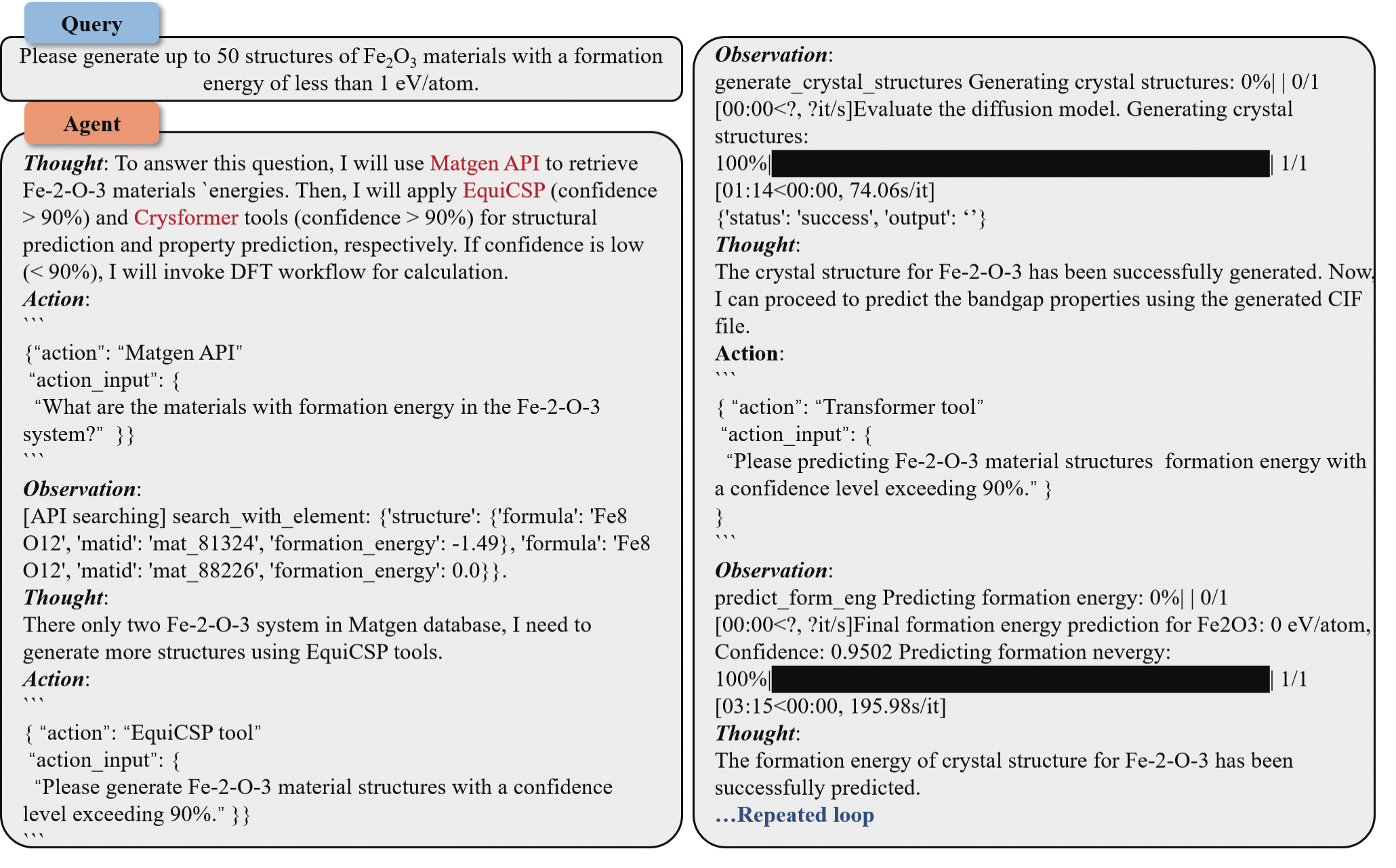}
    \caption{Multi-workflow retrieval-augmented generation for materials informatics.}
    \label{fig:promp}
\end{figure*}

\begin{figure*}[htp]
    \centering
    \includegraphics[width=0.95\textwidth]{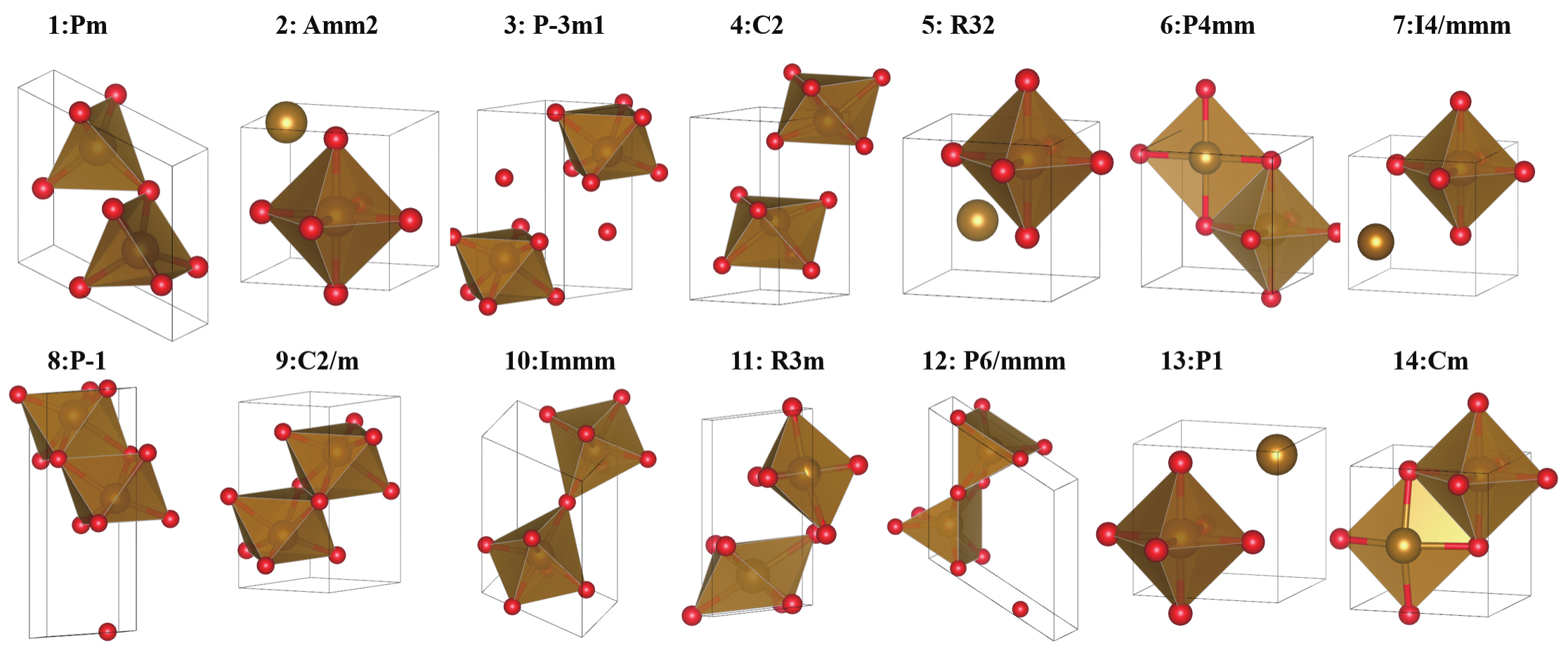}
    \caption{Structural design for Fe$_2$O$_3$: AI-generated material structures optimized via DFT, resulting in 14 valid structures and their corresponding space groups after multiple iterations.}
    \label{fig:val_str}
\end{figure*}

\begin{figure}
    \centering
    \includegraphics[width=0.4\textwidth]{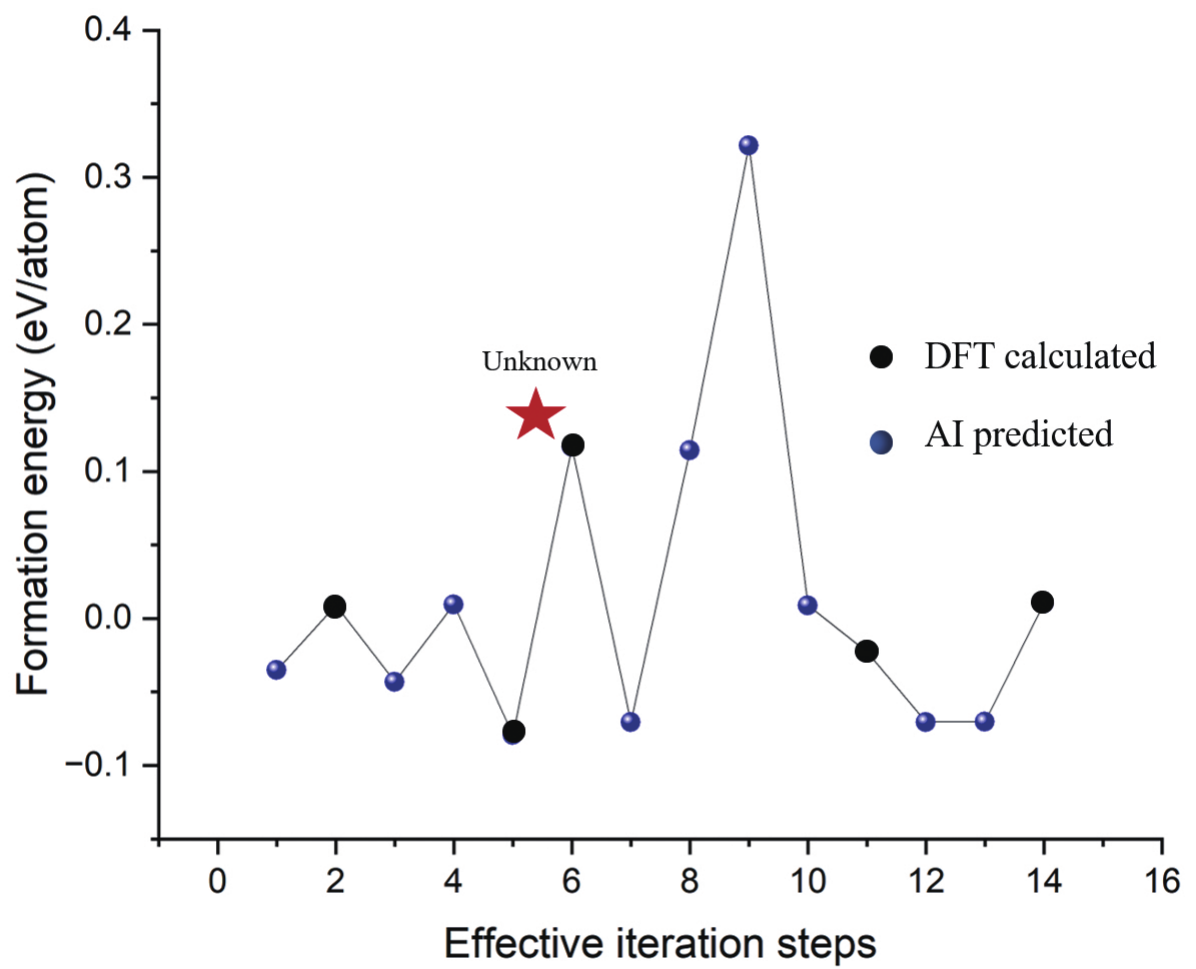}
    \caption{Iterating 50 steps to discover potential novel structures. 
Failed and duplicate material structures are omitted here.}
    \label{fig:iterative}
\end{figure}

Furthermore, we evaluated two sampling strategies for $p(N)$: one based on the pre-trained data distribution and the other using purely random sampling. As shown in Fig.~\ref{fig:dire_thr}b, the pre-trained distribution initially generates a high proportion of valid structures (43\%), benefiting from the learned structural priors of known stable materials. However, this advantage diminishes over time, dropping to 22\%, as the sampling space becomes saturated and redundant structures are filtered out. In contrast, the random sampling strategy maintains a relatively stable valid generation rate around 28\%, due to its broader exploration of the compositional and structural space, despite the lower initial success rate. These results suggest that a hybrid strategy—starting with pre-trained distribution sampling to efficiently generate high-quality structures, followed by random sampling to enhance diversity—offers a more effective approach for valid structure discovery.

We conducted a statistical analysis of approximately 10 million potentially valid material structures generated by the model. As shown in Fig.\ref{fig:dire_gen}a, the chemical element distribution of these structures spans nearly the entire periodic table, encompassing 82 different element types. Most structures have atomic counts below 30, although a significant number of structures contain more than 80 atoms (Fig.\ref{fig:dire_gen}b). Furthermore, the crystal systems of the generated material structures include all seven categories. Among these, the monoclinic system accounts for the largest proportion at 27.2\%, while the triclinic system represents the smallest fraction at 1.2\% (Fig.\ref{fig:dire_gen}c).

Figure~\ref{fig:prop_dist} presents the frequency distributions of the predicted and DFT-calculated values for bandgap and formation energy. Fig~\ref{fig:prop_dist}a and c compare the bandgap distributions, where the AI-predicted values exhibit a narrower range, with a peak around 0-2 eV and a rapid decline for higher values. In contrast, the DFT-calculated distribution demonstrates a smoother decay and broader coverage, particularly for bandgap values exceeding 6~eV. Fig~\ref{fig:prop_dist}b and d analyze formation energy distributions, showing that AI predictions are concentrated between -2 and 4~eV/atom, with a symmetric peak near 1 eV/atom. The DFT results, however, display a significantly wider range, capturing negative values and extending up to 12~eV/atom. These differences highlight the trade-off between the efficiency and generalization of AI models and the comprehensive nature of DFT calculations, emphasizing the necessity of validating AI predictions against high-fidelity DFT data to ensure robustness, particularly for rare or extreme material properties.

\subsection{Coordinate Pattern for LLM-based Material Designing}

In this experiment, we will integrate the previously developed model, Crysformer, to perform DFT calculations on generated material structures. Given the vast compositional space of material structures, we will focus on calculating the structures of potentially stable materials. Specifically, Crysformer will be used to predict formation energies and provide confidence scores. For material structures with low confidence scores, DFT calculations will be performed to ensure accurate evaluations.
 
As illustrated in Fig.~\ref{fig:promp}, we employ an LLM-based agent for materials design, addressing the specific query: "Please generate up to 50 structures of Fe$_2$O$_3$ materials with a formation energy of less than 1 eV/atom." In response, the agent initially retrieves existing structures from the database, identified as mat\_81324 and mat\_88226. To further explore novel Fe$_2$O$_3$ structures, the agent utilizes the EquiCSP tool for structure generation and predicts the formation energies of the generated structures. Considering that potential novel structures may not be included in the training dataset of Crysformer, we assess the model's confidence to decide its applicability. If the model confidence falls below 90\%, numerical simulations are performed using DFT software. The 90\% confidence threshold was chosen based on model calibration and supported by reported DFT formation energy MAEs of 0.081–0.136 eV/atom~\cite{xie2018crystal}.
We conservatively adopt 0.12 eV/atom as the acceptable upper bound for prediction error, and 90\% confidence corresponds to the region where model predictions typically fall within this range.
This iterative process is repeated 50 times. Subsequently, duplicate structures and those with a space group of 1 are filtered out, yielding 14 unique structures with formation energies below 1 eV/atom.
Fig.~\ref{fig:val_str} presents the detailed 3D structures of the 14 designed materials along with their corresponding space group information. 
Fig.~\ref{fig:iterative} presents the formation energies obtained during the iterative process, combining HPC and AI computations. In these 50 iterations, the AI model was applied nine times to predict formation energies. When the model exhibited low confidence, five DFT calculations were performed to ensure reliability. This HPC–AI hybrid approach effectively circumvented nine additional time-consuming DFT computations, thereby enhancing the overall efficiency of the materials design workflow.

\section{Discussion}
\begin{table*}
\centering
\caption{Comparison of HPC-AI Integration patterns.}
\begin{tabular}{lp{4.0cm}p{4.0cm}p{5.0cm}}
\hline
\textbf{pattern} & \textbf{Generalization Capacity} & \textbf{Computational Efficiency} & \textbf{Applicability to Real Scenarios} \\ \hline
\textbf{Surrogate} & Limited by data; struggles with unseen conditions. & High efficiency by bypassing HPC. & Requires confidence evaluation for reliability. \\ \hline
\textbf{Directive} & Moderate; supported by physics-based HPC. & Efficient in optimization; HPC-intensive. & Strong potential for optimization-centric problems. \\ \hline
\textbf{Coordinate} & High; combines AI and HPC for robust predictions. & Balanced; AI reduces HPC workload. & Highly versatile; ideal for iterative workflows. \\ \hline
\end{tabular}
\label{tab:hpc_ai_modes}
\end{table*}

The integration of HPC and AI into three distinct patterns—surrogate, directive, and coordinate—can be systematically analyzed across three critical dimensions: generalization capacity, computational efficiency, and usability to real scientific scenarios as show in Table~\ref{tab:hpc_ai_modes}. 

\subsection{Generalization Capacity}
Surrogate pattern leverages data-driven AI models to approximate HPC computations. Its generalization capacity is limited by the scope and quality of the training dataset. Models trained on insufficiently diverse or under-representative data may struggle to accurately predict results for novel or complex scientific problems. This limitation highlights the inherent trade-off between computational simplicity and predictive robustness. While directive pattern also depends on AI models, its reliance on physics-based HPC computations provides a safeguard against AI generalization limitations. The combination of data-driven insights with physical principles enhances reliability but does not completely overcome the AI’s challenges in extrapolating beyond the training data. Coordinate pattern addresses the generalization problem by integrating AI predictions with high-confidence HPC validation. The iterative feedback between AI and HPC reduces the risk of overfitting or extrapolation errors, thus enhancing the generalization capacity compared to standalone AI or surrogate approaches.

\subsection{Computational Efficiency}
Surrogate patterns excel in computational efficiency by completely replacing resource-intensive HPC calculations with AI predictions. This efficiency enables rapid exploration of large parameter spaces or complex scenarios that would otherwise be computationally prohibitive. However, the reliance on a pre-trained model means the efficiency is front-loaded and may degrade in scenarios requiring frequent retraining for new conditions. The directive pattern achieves a balance between efficiency and accuracy by using AI to streamline the parameter optimization phase. Rather than explicitly shrinking the parameter space, AI-guided optimization focuses computational effort on regions more likely to yield optimal or valid outcomes, effectively narrowing the explored search space. For instance, in our structure generation task, the learned generative model prioritizes high-quality candidates by leveraging prior distributions, thereby reducing the number of expensive but unproductive evaluations.
However, the final stages of HPC calculations, such as solving large-scale differential equations or conducting DFT computations, remain resource-intensive, making this pattern less efficient than surrogate approaches. The cooperative nature of this pattern combines AI’s computational speed with HPC’s precision. By offloading simpler calculations to AI and reserving HPC for critical validation steps, the overall computational load is reduced. This selective allocation of resources results in a significant efficiency gain compared to directive pattern, while retaining the accuracy benefits of HPC.

\subsection{Usability to Real Scientific Scenarios}
The usability of surrogate patterns in real scientific problems is constrained by their confidence and reliability. In domains where uncertainty quantification is critical, such as materials discovery or drug design, surrogate patterns must incorporate mechanisms to evaluate and report prediction confidence. Without such mechanisms, their deployment in high-stakes scenarios remains limited. The directive pattern offers strong applicability in scientific contexts requiring iterative optimization, such as the search for optimal material properties or parameter tuning in large-scale simulations. The combination of AI-guided exploration and HPC’s rigorous validation makes this pattern particularly suitable for the problems demanding both exploration and precision. Coordinate pattern is the most versatile for real scientific applications. By integrating AI and HPC in a collaborative workflow, it ensures both speed and accuracy. For instance, in materials science, AI predictions can guide initial exploration, while HPC methods like DFT validate and refine results. Moreover, the validated outputs can serve as enhanced training data for AI, creating a feedback loop that continuously improves both efficiency and scientific insight.

\section{Related work}

\subsection{Surrogate pattern}
In the domain of materials science, surrogate AI models have emerged as powerful tools for accelerating the prediction of electronic structures, properties, and optimized molecular configurations. Unlike traditional HPC-based simulations that rely on first-principles methods such as DFT or on large-scale molecular dynamics (MD) simulations, surrogate patterns leverage data-driven approaches to approximate the underlying physics at significantly reduced computational cost. For instance, electronic structure prediction AI models such as ChargeE3Net~\cite{koker2024higher} and Hamiltonian estimation frameworks like DeepH~\cite{li2022deep} have been developed to predict electronic charge distributions and Hamiltonians, respectively, from representative training data without the need to solve the full set of quantum mechanical equations at runtime. Additionally, property prediction AI models including CrystalNet~\cite{chen2022improving}, ALIGNN~\cite{choudhary2021atomistic} and Matformer~\cite{Yan2022PeriodicGT} have shown remarkable success in estimating material properties (e.g., formation energies, band gaps) based on crystal graphs or chemical compositions. Molecular structure optimization AI models, exemplified by DPA2~\cite{zhang2023dpa}, directly suggest low-energy configurations of molecules or crystals, circumventing exhaustive searches.

A common characteristic of these surrogate AI models is that they are trained on either experimentally measured data or on computational datasets previously generated by DFT and other HPC-based simulations. While this approach substantially reduces computational overhead and allows for rapid inference, the generalization capability of surrogate AI models is often limited by the scope and quality of the training data. Consequently, their applicability to materials previously unseen or outside the distribution of the training set can be compromised. Recent research efforts focus on improving AI model robustness through uncertainty quantification, domain adaptation, and physics-informed neural networks, ensuring that these surrogates remain reliable tools for materials discovery and design.

\subsection{Directive pattern}
Directive pattern integrates AI models as guides within HPC workflows, leveraging data-driven insights to enhance the efficiency and accuracy of physically rigorous simulations. In materials science, one prominent example is DeepMD~\cite{jia2020pushing}, which employs neural networks to learn atomic force fields from DFT reference calculations. By accurately capturing interatomic potentials, DeepMD can direct classical MD simulations towards physically meaningful trajectories with fewer computations. This approach improves the search and exploration of stable crystal structures, reaction pathways, or phase diagrams by mitigating the inefficiencies of randomly sampling vast configuration spaces.

Similar methods, which we collectively denote as ``AI-augmented HPC frameworks'', have explored a range of strategies to guide simulations, such as employing Bayesian optimization to target regions of chemical space with high potential for desired properties~\cite{jablonka2021bias}. Although directive AI model still depends on HPC for final validation and refinement, the interplay between AI-guided exploration and physics-based solvers results in more focused and informative searches. This synergy helps to reduce the cost of large-scale computations while maintaining scientific rigor, ultimately accelerating the materials design cycle.

\subsection{Coordinate pattern}
Coordinate pattern represents a cooperative paradigm where HPC and AI models iteratively inform and improve one another. This pattern is often realized through reinforcement learning, active learning, or agent-based methods that dynamically update both AI predictive models and simulation parameters as new information is obtained. For example, frameworks like LLaMP~\cite{chiang2024llamp}, ChatMOF~\cite{kang2024chatmof}, ChemCrow~\cite{m2024augmenting} (an AI-driven pipeline for materials exploration) employ an agent-based approach. 
%where predictions from AI models are continually validated and refined using HPC simulations~\cite{chiang2024llamp}
The validated outcomes, in turn, serve as new training data that enhance the AI models' accuracy and generalization.

In practice, this coordinated approach creates a closed-loop system in which HPC computations and AI predictions form a feedback cycle. AI quickly proposes candidate materials or configurations, HPC tests these candidates at high fidelity, and the results are fed back into the AI models. Over successive iterations, this coordination not only improves the quality of predictions but also reduces the computational load compared to purely brute-force HPC simulations. Such reciprocal refinement is particularly promising for discovering novel materials with tailored properties, enabling more efficient and directed searches of the vast chemical and configurational spaces inherent in materials science.

\section{Conclusion}
In this work, we have investigated the coupling of HPC and AI within the context of materials science, proposing a coupling methodology that encompasses three distinct patterns: surrogate, directive, and coordinate. The surrogate pattern leverages data-driven AI models trained on experimental or DFT-computed data to bypass expensive HPC calculations, thereby accelerating material property predictions and structural optimizations. The directive pattern guides HPC workflows through AI-driven force field fitting or targeted search, balancing the accuracy of physics-based simulations with the efficiency afforded by machine learning. Finally, the coordinate pattern integrates HPC, AI and third-party intelligent roles in a dynamic, closed-loop feedback system, leveraging reinforcement learning, active learning, or agent-based methods to iteratively refine both AI predictive models and simulation parameters.

Our exploration highlights significant performance gains and methodological advancements. These include fast inference and broad usability in surrogate AI models, targeted parameter explorations and improved resource utilization in directive patterns, and enhanced robustness and adaptability in coordinate patterns. By presenting concrete implementations, ranging from materials property predictors such as Crysformer, as well as integration frameworks like EquiCSP and LLM-based Agent, we offer insights into successful HPC-AI coupling.

These interaction patterns, though exemplified in the materials science domain, possess the flexibility to be extended to other scientific fields. Their underlying principles, combining computational rigor with intelligent guidance, can serve as a blueprint for similar HPC-AI collaborations aimed at unraveling complex, high-dimensional problem spaces. As the roles of AI in HPC continue to evolve, the approaches outlined here provide valuable strategies for accelerating discovery, enhancing simulation fidelity, and ultimately expanding the horizons of scientific inquiry.  

\begin{acks}
This research was supported by the Guangdong Provincial Key Area R\&D Program (Grant No. 2024B0101040005).
\end{acks}

\bibliographystyle{SageH}  %{named}
\bibliography{reference}

%\begin{thebibliography}{99}
%\bibitem[Kopka and Daly(2003)]{R1}
%Kopka~H and Daly~PW (2003) \textit{A Guide to \LaTeX}, 4th~edn.
%Addison-Wesley.

%\bibitem[Lamport(1994)]{R2}
%Lamport~L (1994) \textit{\LaTeX: a Document Preparation System},
%2nd~edn. Addison-Wesley.

%\bibitem[Mittelbach and Goossens(2004)]{R3}
%Mittelbach~F and Goossens~M (2004) \textit{The \LaTeX\ Companion},
%2nd~edn. Addison-Wesley.

%\end{thebibliography}

\end{document}